\documentclass{aastex}
\usepackage{spr-astr-addons}
\usepackage{url}

\RequirePackage{color}

\newcommand{\emaila}{tei.naulak@uohyd.ac.in}
\newcommand{\emailb}{sureshpk@uohyd.ac.in}

\begin{document}

\title{BB mode  angular power spectrum of CMB  from massive gravity }

\author{N. Malsawmtluangi\altaffilmark{1}} \and \author{P.K. Suresh}
\affil{\emaila \\ \emailb \\ School of Physics, University of Hyderabad,\\
Hyderabad-500 046, India.}

\altaffiltext{1}{Department of Physics, Government Kolasib College, Kolasib-796081, India}

\begin{abstract}
The  BB-mode correlation angular power spectrum of CMB is  studied for primordial massive gravitational waves  for several inflation models. The   comparative  study of the  angular power spectrum  with  the joint BICEP2/Keck Array and Planck  data  suggests  further constraint on the lower and upper bounds on the mass of primordial gravitons. Assuming a modified dispersion relation, the mass of primordial graviton is also calculated. The resulting constraint also agrees with other theoretical estimates.
\end{abstract}


\section{Introduction}
The force of gravity is believed to be mediated by a spin-2 particle called graviton which is  commonly considered to be massless, thus travelling with the speed of light according to the theory of general relativity. However, starting   with the idea of a spin-2 particle with non-zero mass, several approaches have been taken to introduce mass to  graviton \citep{FP, vDVZ, zak, BD,  vainshtein, GS}. Endowing graviton with mass leads to extra degrees of freedom which do not decouple as  graviton mass approaches to zero such that the general relativistic (GR) case cannot be recovered \citep{vDVZ, zak}. Some of the approaches to massive gravity suffer from pathologies like the presence of ghost mode \citep{BD}, discontinuity when the mass approaches to zero limiting case and so on \citep{vainshtein}, and several theories have been proposed to fix these problems and also to formulate a consistent theory of massive gravity \citep{GS, Rubakov, Dubovsky, dRGT, dRGT2, Sgr}.
At the same time, there have been several attempts to estimate the mass of graviton from astrophysical sources as well as  for  primordial gravitational waves (GWs)  \citep{m1, m2, m3, m4, m5, m6, m7}. It is believed that if the mass of the graviton is comparable to the Hubble parameter, then the massive graviton would condensate to form effective negative pressure stress energy at cosmological distances which would provide  a repulsive effect thus leading to  late  time cosmic acceleration, thereby  suggesting that the massive gravitons could be responsible  for  the current accelerating phase of the universe  instead of dark energy. There are also studies that  propose that massive gravitons would  comprise of cold dark matter as well \citep{cdm}. 
  
In this paper, we consider the particular Lorentz-violating massive gravity theory in which the Lorentz invariance is violated through spontaneous symmetry breaking caused by the presence of background Goldstone fields which leads to the modification of the dispersion relation. The Goldstone fields are set to their vacuum values and the resulting mass parameters are fine-tuned relative to each other in such a way that the pathologies are absent, and the scalar and vector modes behave exactly like those in the general relativistic case. Hence,  the modification of the gravity  comes only from the tensor modes and the dispersion relation of gravitational waves  acquires an effective mass and is relativistic \citep{Rubakov, Dubovsky}. According to this theory, the bound on the primordial graviton mass is obtained from the exponential decay in the Yukawa potential, putting the upper bound for the graviton mass to be $\le 10^{-30}$ eV at the Compton wavelength of $\lambda_g > 10^{20}$ km \citep{Dubovsky, cmbmg2}. The lower bound for graviton mass has been proposed to be $> 10^{-29}$ $cm^{-1}$ ($ \equiv 1.239 \times 10^{-32}$ eV)  \citep{mgtensor}. The minimum for the mass of graviton in the de Sitter spacetime has also been set by the Higuchi bound as $m_{gw}^2 \ge 2H^2$, where $H$ is the Hubble parameter \citep{hig}. The small mass of graviton is expected to have an effect on the temperature anisotropy and polarization spectra of the cosmic microwave background (CMB)  \citep{cmbmg2, cmbmg1}. The  imprint of  primordial gravitational waves   on CMB anisotropy  can be observed through the angular power spectrum of CMB in the form of  B-mode polarization \citep{marc, marc2, baskaran, lpg}. The  observation of B-mode polarization on CMB can  not only verify the theory of inflation itself but would also help in constraining the many inflation models  \citep{jmart1, jmart2, jmart3}. The detection of B-mode polarization of CMB or the primordial GW itself  would be able to provide a clear bound on the mass of primordial graviton. Hence, in this paper, we  study the effect of the primordial massive GWs on the BB mode correlation angular power spectrum  of CMB for various inflation models and  the results are compared with  the  2015 BICEP2/Keck Array  and Planck  collaboration data \citep{bkp} and therefore try to  obtain a constraint on the  mass of  primordial gravitons.

\section{Massive gravitational waves}
Action for massive gravity can be written in terms  of the Einstein-Hilbert action and the Goldstone action as  \citep{Rubakov, Dubovsky},
\begin{eqnarray}\label{action}
S &=& S_{EH} + S_{G}, \nonumber \\ 
 &=&\int d^4x \sqrt{-g}[-m_{pl}^2R + \Lambda^4F(Z^{ij})],
\end{eqnarray}
where $\Lambda$ characterizes the cutoff energy scale for low energy effective theory. F is a function of the Goldstone field, metric components and  its derivatives. The second term in the  above action leads to violation of  the Lorentz symmetry.   It is assumed that ordinary matter field is minimally coupled to the metric.

Action depends on the Goldstone field derivatives through the argument $Z^{ij}$ which can be obtained with the help of the  following expressions,
\begin{eqnarray}
Z^{ij} &=& X^{\gamma}W^{ij}, \nonumber \\
X &=& \Lambda^{-4}g^{\mu \nu} \partial_{\mu}\Phi^0\partial_{\nu}\Phi^0, \nonumber \\
W^{ij} &=& \Lambda^{-4}g^{\mu \nu} \partial_{\mu}\Phi^i\partial_{\nu}\Phi^j - \frac{V^iV^j}{X}, \nonumber \\
V^i &=& \Lambda^{-4}g^{\mu \nu} \partial_{\mu}\Phi^0\partial_{\nu}\Phi^i,
\end{eqnarray}
where  $\Phi^0 (x)$, $\Phi^i (x)$, ($i = 1,2,3$) are the four scalar fields and $\gamma$ is considered as a constant free parameter.

For  Eq.\ref{action}, the vacuum solutions corresponding to  the  flat Friedmann-Lemaitre-Robertson-Walker (FLRW) metric are obtained after setting the Goldstone fields to their vacuum values and can be written as
\begin{eqnarray}\label{ug}
g_{\mu \nu} &=& a^2 \eta_{\mu \nu}, \nonumber \\
\Phi^0 &=& \Lambda^2t, \\
\Phi^i &=& \Lambda^2 x^i. \nonumber
\end{eqnarray}
where $a $ is the scale factor  for the FLRW metric and  $\eta_{\mu \nu}$ is the flat space metric. 

The metric  $g_{\mu \nu}$  with perturbations can be written as
\begin{equation}\label{met}
g_{\mu \nu} = a^2\eta_{\mu \nu} + \delta g_{\mu \nu},
\end{equation}
where  the metric perturbations $\delta g_{\mu \nu}$ are taken   after the spontaneous Lorentz symmetry breaking.

The metric perturbations can be decomposed as,
\begin{eqnarray}
\delta g_{00} &=& 2a^2 \varphi, \nonumber \\
\delta g_{0i} &=& a^2 (N_i - \partial_i A), \\
\delta g_{ij} &=& a^2 [-h_{ij} - \partial_iQ_j - \partial_jQ_i + 2(\psi \delta_{ij} - \partial_i \partial_j E)], \nonumber
\end{eqnarray}
where  $\varphi$, $\psi$, A and E are scalar fields, $N_i$ and $Q_i$ are transverse vector fields and $h_{ij}$ is the transverse-traceless tensor perturbation.

 By expanding $\sqrt{-g + \delta g}$, $X(g + \delta g)$, $V^i(g + \delta g)$, $W^{ij}(g + \delta g)$ in  Eq.\ref{ug} and using Eq.\ref{action} we get the Lagrangian as
\begin{eqnarray}\label{lm}
L_m &=& \frac{m_{pl}^2}{2}(m_0^2h_{00}h_{00} + 2m_1^2h_{0i}h_{0i} - m_2^2h_{ij}h_{ij} \nonumber \\
 && + m_3^2h_{ii}h_{jj} - 2m_4^2h_{00}h_{ii}),
\end{eqnarray}
where the mass parameters  are given by \citep{mvp},
\begin{eqnarray}
m_0^2 &=& \frac{\Lambda^4}{m_{pl}^2}[XF_X + 2X^2 F_{XX}], \nonumber \\
m_1^2 &=& \frac{2 \Lambda^4}{m_{pl}^2}[-XF_X -WF_W + \frac{1}{2}XWF_{VV}], \nonumber \\
m_2^2 &=& \frac{2 \Lambda^4}{m_{pl}^2}[WF_W - 2W^2F_{WW2}], 
\end{eqnarray}
\begin{eqnarray}
m_3^2 &=& \frac{\Lambda^4}{m_{pl}^2}[WF_W + 2W^2F_{WW1}], \nonumber \\
m_4^2 &=& -\frac{\Lambda^4}{m_{pl}^2}[XF_X + 2XWF_{XW}], \nonumber
\end{eqnarray}
where
\begin{eqnarray}
W &=& -1/3 \delta_{ij} W^{ij}, \nonumber \\
\frac{\partial F}{\partial X} &=& F_X, \nonumber \\
\frac{\partial^2 F}{\partial X^2} &=& F_{XX}, \nonumber \\
\frac{\partial F}{\partial W^{ij}} &=& F_W\delta_{ij}, \\
\frac{\partial^2 F}{\partial V^i \partial V^j} &=& F_{VV}\delta_{ij}, \nonumber \\
\frac{\partial^2 F}{\partial W^{ij} \partial W^{kl}} &=& F_{WW1}\delta_{ij}\delta_{kl} + F_{WW2}(\delta_{ik}\delta_{jl} +\delta_{ij}\delta_{jk}), \nonumber \\
\frac{\partial^2 F}{\partial X \partial W^{ij}} &=& F_{XW}\delta_{ij}. \nonumber
\end{eqnarray}
For the flat cosmological solutions, $X=a^{-2}\Phi'^2$, $V^i=0$, $W^{ij}=-a^{-2}\delta^{ij}$. The Einstein field equations for Eq.\ref{action} with the scalar fields in the unitary gauge Eq.\ref{ug} and the metric Eq.\ref{met} then reduce to the following relations,
\begin{eqnarray}
\left(\frac{a'}{a}\right)^2 &=& \frac{a^2}{m_{pl}^2} [\rho_m + \Lambda^4 (2XF_X-F)] \\
2\frac{a''}{a}- \left(\frac{a'}{a}\right)^2 &=& -\frac{a^2}{m_{pl}^2}[p_m + \Lambda^4(2WF_W+F)]
\end{eqnarray}
where $\rho_m$ and $p_m$ are the energy density and pressure respectively for ordinary matter and the equation of motion of the $\Phi^0$ field,
\begin{equation}
\partial_0(a^3 F_X X^{1/2})=0.
\end{equation}
Prime here denotes derivative with respect to conformal time $\eta$. Apart from some constraints which arise from the requirement that the model is free of ghosts and strong coupling problems, the function $F$ is quite arbitrary. Specific restrictions on the function $F$ are discussed in detail in \citep{Dubovsky, cdm} where the existence of a wide class of functions with graviton masses are demonstrated.

The mass parameters are carefully fine tuned relative to each other. The fine tuning relations between the mass parameters characterize certain regions in the mass parameter space so that in these regions, the theory is free of pathologies, and the theory is described by a consistent low-energy effective theory with strong coupling scale $\Lambda \sim (m m_{pl})^{1/2}$ which implies a ghost-free scenario  \citep{Rubakov, Dubovsky}. The mass parameter $m_2$ represents the mass of the graviton which arises from the modification in the tensor sector in which there are two massive spin-2 propagating degrees of freedom. The vector and scalar perturbations behave similarly as in the general relativity case.

The perturbed metric for  a flat FLRW universe can  be written as
\begin{equation}
dS^2 = a^2(\eta) [-d\eta^2 + (\delta_{ij} + h_{ij}) dx^i dx^j],
\end{equation}
here $\delta_{ij}$ is the flat space metric 
and $\eta$ is the conformal time defined by $d\eta = \frac{dt}{a}$.

The dynamical equation of motion for massive gravitational waves can be written as
\begin{equation}\label{emij}
h_{ij}^{(m) \prime \prime}(\eta) + 2Hh_{ij}^{(m)\prime}(\eta) + k^2h_{ij}^{(m)}(\eta) + a^2 m_{gw}^2 h_{ij}^{(m)}(\eta)  = 0,
\end{equation}
where $m_{gw} \equiv m_2$ is the mass of the graviton  and  $H = \frac{a'}{a^2}$ is the Hubble parameter.

The massive tensor perturbation $h_{ij}^{(m)}$ can be expanded in the Fourier space as
\begin{eqnarray}\label{fmm}
h_{ij}^{(m)} (\textbf{x}, \eta) &=& \frac{D}{(2\pi)^{\frac{3}{2}}} \int^{\infty}_{-\infty} \frac{d^3\textbf{k}}{\sqrt{2E_k}} \nonumber \\
&& [h_{k}^{(m)(p)} (\eta) c_{k}^{(m)(p)} \varepsilon_{ij}^{(m)(p)} (\textbf{k}) e^{i\textbf{k}.\textbf{x}} \\
&& + h_{k}^{(m)(p) \ast} (\eta) c_{k}^{(m)(p) \dagger} \varepsilon_{ij}^{(m)(p) \ast} (\textbf{k}) e^{-i\textbf{k}.\textbf{x}}] \nonumber ,
\end{eqnarray}
where $D = \sqrt{16\pi}l_{pl}$ is the normalization constant, $l_{pl}$
 is  the Planck length, $E_k$ is the energy of the mode ,$(p)$ is the polarization index and the superscript $(m)$ stands for  the massive tensor perturbation.

The two polarization states $\varepsilon_{ij}^{(p)}$, $p=1,2$ are symmetric and transverse-traceless and satisfy the conditions
\[\varepsilon_{ij}^{(p)}\delta^{ij}=0, ~~\varepsilon_{ij}^{(p)}k^i=0,\] \[\varepsilon_{ij}^{(p)}\varepsilon^{(p') ij} = 2\delta_{pp'}, ~~\varepsilon_{ij}^{(p)}(\textbf{-k})=\varepsilon_{ij}^{(p)}(\textbf{k}).\]
These polarizations are linear and are called the plus $(+)$ polarization and cross $(\times)$ polarization. 

The creation and annihilation operators $c_k^{(p) \dagger}$ and $c_k^{(p)}$ satisfy the  following relations
\begin{eqnarray}
\left[c_k^{(p)},c_{k'}^{(p') \dagger}\right]&=&\delta_{pp'}\delta^3(k-k'),\\
\left[c_k^{(p)},c_{k'}^{(p')}\right]&=&\left[c_k^{(p) \dagger},c_{k'}^{(p') \dagger}\right]=0.
\end{eqnarray}

Using  Eq.\ref{fmm}  in Eq.\ref{emij}, we get
\begin{equation}\label{emk}
h_k^{(m)\prime \prime} (\eta) + 2H h_k^{(m)\prime} (\eta) + (k^2 + a^2m_{gw}^2) h_k^{(m)} (\eta) = 0.
\end{equation}
Hereafter we drop  the polarization index $(p)$ and the index $(m)$ for notational convenience.

The mode function  can be taken in the following form
\begin{equation}\label{mf}
\mu_k (\eta) = a (\eta) h_k (\eta).
\end{equation}
Using  Eq.\ref{mf}  in Eq.\ref{emk} we get 
\begin{equation}\label{mfm}
\mu_k^{\prime \prime} + \left(k^2 + a^2 m_{gw}^2 -\frac{a''}{a} \right) \mu_k = 0.
\end{equation}
The dispersion relation can be written as \citep{gum}
\begin{equation}\label{dr}
\frac{k^2}{a^2} + m_{gw}^2 = w^2,
\end{equation}
where $w$ is known as the effective frequency. 

For the adiabatic vacuum, Eq.\ref{emk} has the solution
\begin{equation}\label{av}
h_k (\eta) \propto e^{-iwa\eta}.
\end{equation}

For super horizon modes, $w^2 \ll H^2$, the tensor amplitudes are frozen and the mode stays outside the horizon and its absolute value is
\begin{equation}\label{sb}
|h_k| = \mathcal{A}_{ex}(k), ~~~~\eta < \eta_{k},
\end{equation}
where $\mathcal{A}_{ex} (k) = \frac{H_{ex}}{m_{pl} k^{3/2}}$, is the amplitude of the mode at the time of its generation and  $H_{ex}$ is the expansion rate at the time of horizon exit during inflation,  $\eta_{k}$ is the time of horizon re-entry and $m_{pl}$ is the reduced Planck mass.

On horizon crossing, $w^2 \simeq H^2$. Assuming that the horizon re-entry takes place sufficiently rapidly, i.e., $\eta \simeq \eta_{k}$, then Eq.\ref{av} can be rewritten as 
\begin{equation}\label{hre}
h_k (\eta) = \frac{\mathcal{C}(k)}{\sqrt{w_{k} a_{k}^3}} e^{-iwa\eta}, ~~~~\eta \simeq \eta_{k},
\end{equation}
where $w_{k} \equiv w(\eta_{k}) = H_{k}$ indicates horizon re-entry.

On horizon re-entry, the frequency of the wave mode becomes higher than the rate of cosmic expansion, $w^2 \gg H^2$, such a mode is called sub-horizon mode. Its solution is given by Eq.\ref{av}:
\begin{equation}\label{sp}
h_k (\eta) = \frac{\mathcal{C}(k)}{\sqrt{w(\eta) a^3(\eta)}} e^{-iwa\eta}, ~~~~\eta > \eta_{k},
\end{equation}
where $\mathcal{C}(k)$ is a constant of integration.

Using Eq.\ref{sb}, Eq.\ref{hre} and Eq.\ref{sp}, we get
\begin{equation}\label{sm}
\frac{|h_k(\eta)|}{\mathcal{A}_{ex}(k)} = \sqrt{\frac{w_{k}}{w(\eta)}\frac{a_{k}^3}{a^3(\eta)}}, ~~~~\eta > \eta_{k}.
\end{equation}
Replacing $w$ by $k/a$ and $\eta_{k}$ by $\eta_{k}^{GR}$, $GR$ indicating the massless case, we get the corresponding solution in the massless case as
\begin{equation}\label{sml}
\frac{|h_k^{GR}(\eta)|}{\mathcal{A}_{ex}(k)} = \frac{a_{k}^{GR}}{a(\eta)}, ~~~~\eta > \eta_{k}^{GR}.
\end{equation}
The two-point correlation function  for the massive gravitational waves can be written as
\begin{equation}
P(w_0) \equiv \frac{d}{d \ln w_0} \langle 0| h_{ij} (\textbf{x},\eta) h^{ij} (\textbf{x},\eta) |0 \rangle,
\end{equation}
where 
\begin{equation}
\langle 0| h_{ij} (\textbf{x},\eta) h^{ij} (\textbf{x},\eta) |0 \rangle = \frac{D^2}{2\pi^2} \int^{\infty}_0 k^2 |h_k (\eta)|^2 \frac{dk}{k}.
\end{equation}
Therefore one gets
\begin{equation}
P(w_0) = \frac{w_0^2}{w_0^2 - m_{gw,0}^2} \frac{2k^3}{\pi^2}|h_k(\eta_0)|^2,
\end{equation}
where 
\begin{eqnarray}
k = a_0\sqrt{w_0^2 - m_{gw,0}^2}, \label{jk} \\
\frac{d}{d \ln w_0}\left(\frac{dk}{k}\right) = \frac{w_0^2}{w_0^2 - m_{gw,0}^2}.
\end{eqnarray}
where the subscript '0' represents evaluation at present time. Using Eq.\ref{sm}, the  power spectrum for the massive gravitational waves  is obtained as
\begin{eqnarray}\label{pm}
P(w_0) &=& \frac{2k^3}{\pi^2} \mathcal{A}^2(k) \left(\frac{k'a_{k}}{ka_0}\right)^2 \frac{w_{k}a_{k}}{w_0a_0} \nonumber \\
	&=& \left(\frac{k'a_{k}}{ka_0}\right)^2 \frac{w_{k}a_{k}}{w_0a_0} P(k),
\end{eqnarray}
where $k' = a_0w_0$ and $P(k) = \frac{2k^3}{\pi^2} \mathcal{A}^2(k)$ is known as the primordial power spectrum.

Using Eq.\ref{sml}, the power spectrum for the massless case can be written as
\begin{equation}\label{pml}
P_{GR} (w_0) = \left(\frac{a_{k'}^{GR}}{a_0} \right)^2 P(k').
\end{equation}
By taking the ratio of Eq.\ref{pm} to Eq.\ref{pml}, we obtain
\begin{eqnarray}\label{ps}
\frac{P(w_0)}{P_{GR}(w_0)} &=& \frac{P(k)}{P(k')} \left(\frac{k'a_k}{ka_{k'}^{GR}}\right)^2 \frac{w_ka_k}{w_0a_0} \nonumber \\
	&=& \frac{P(k)}{P(k')} S^2(w_0),
\end{eqnarray}
where the enhancement factor  $ S(w_0)$ can be written as
\begin{equation}\label{ef}
S(w_0) = \frac{k'a_k}{ka_{k'}^{GR}} \sqrt{\frac{w_ka_k}{w_0a_0}}.
\end{equation}
 The dispersion relation  at the time of horizon re-entry is
\begin{equation}
w_k \simeq m_{gw}(\eta_{k}).
\end{equation}

The cosmic expansion rate is comparable to the effective mass of the gravitational waves when all modes re-enter the horizon simultaneously, then
\[H(\eta_{k}) \simeq m_{gw}(\eta_{k}).\]

\noindent
Therefore, we have   $\eta_{k} \simeq \eta_{hc}$, $a_{k} \simeq a_{hc}$, $H_{k} \simeq H$ and $w_{hc} \simeq m_{gw}(\eta_{hc}) = \frac{k_{hc}}{a_{hc}}$.

By considering  the mass term which dominates the frequency modes till present time, we get
\begin{eqnarray}
w_0 \simeq m_{gw,0} = \frac{k_0}{a_0}, \nonumber \\
k' \simeq k_0. \nonumber
\end{eqnarray}
  For long wavelength modes,  the enhancement factor becomes \citep{gum}
\begin{equation}\label{eh}
S(w_0) \simeq \frac{a_{hc}}{a_{k_0}^{GR}}\sqrt{\frac{k_{hc}}{k_0}}\left(\frac{w_0^2}{m_{gw,0}^2}-1\right)^{-\frac{1}{2}}.
\end{equation}

The massive short wavelength modes behave almost similar to their massless counterparts and hence, are not considered here.

\section{Inflation}
In the simplest inflationary scenario, the exponential expansion is driven by a canonical scalar field called the inflaton. In the slow roll scenario, the inflaton slowly rolls down its potential which is almost flat and as long as the slow-roll conditions are satisfied, inflation continues. In most models of slow-roll inflation, the inflation process ends by violation of slow-roll condition which is usually followed by decay of the inflaton and reheating. There are also several models in which the inflaton need not necessarily decay and reheating occurs via some other process.

The equation of motion for the inflaton with effective potential $V$ can be written as
\begin{equation}
\ddot{\phi} + 3H \dot{\phi} + V'(\phi) =0,
\end{equation}
where  the Hubble parameter $H$     
 is determined by the energy density of the scalar field,
 \[ \rho_{\phi} = \frac{\dot{\phi}^2}{2}+V,\] so that the Friedmann equation can be written as
\begin{equation}\label{fdmeq}
H^2 = \frac{1}{3 m^2_{pl}}\left( \frac{1}{2} \dot{\phi}^2 + V(\phi)\right).
\end{equation}
In the slow-roll limit, 
the Hubble parameter and the inflaton potential are related as
\begin{equation}
H^2 \simeq \frac{V}{3m^2_{pl}}.
\end{equation}
 The slow-roll condition is characterized in terms of  the slow-roll parameters  defined in terms of the inflaton potential  and its derivatives as  follows
\begin{eqnarray}\label{sl}
\epsilon &\equiv & \frac{m_{pl}^2}{2}\left(\frac{V'}{V}\right)^2,  \nonumber \\ 
\eta &\equiv &  m^2_{pl}\left(\frac{V''}{V}\right).
\end{eqnarray}
Slow roll conditions demand that $\epsilon$, $\eta \ll 1$. As long as the slow-roll conditions are satisfied, the process of exponential expansion continues and the slow-roll approximation can be used to study the fluctuations generated during inflation. Inflation ends as soon as the slow-roll conditions are violated.

The duration of inflation is characterized by the e-folding number $N$, which can be written in terms of the potential as,
\begin{equation}
N \simeq \frac{1}{m_{pl}^2} \int_{\phi_{end}}^{\phi} \frac{V}{V'}d\phi.
\end{equation}
Throughout this paper, we use $N=60$.

 There are several inflation models and most of them predict the existence of an almost scale invariant tensor perturbations or primordial gravitational waves.
The tensor spectral index describes the deviation of the tensor perturbations from scale invariance and can be written in terms of the parameter $\epsilon$ as
\begin{equation}\label{nT}
n_T = -2\epsilon
\end{equation}

The strength of the tensor fluctuations can be measured with respect to that of the scalar fluctuations and can be realized through the parameter $r$, called the tensor-to-scalar ratio as
\begin{equation}\label{r}
r \equiv \frac{P_T (k)}{P_S (k)} \simeq 16\epsilon,
\end{equation}
where $P_T$ and $P_S$ are the power spectra of the tensor and scalar perturbations respectively,
\begin{eqnarray}
P_T &=& \frac{2}{3\pi^2 m_{pl}^4} V, \\
P_S &=& \frac{1}{12\pi^2 m_{pl}^6} \frac{V^3}{V'^2},
\end{eqnarray}
where $V$ is evaluated at the time when the mode with the wave number $k$ crosses the horizon.

From Eq.\ref{nT} and Eq.\ref{r}, one can see that both $n_T$ and $r$ are determined by the equation of state during inflation, hence these can be very helpful in understanding the dynamics of the early universe and can also help in distinguishing the inflation models. The scalar spectral index $n_s$, on the other hand, must be sufficiently close to scale invariance.

\subsection{Inflation models}
In this work, we consider  the single field slow-roll inflation models for which the corresponding tensor-to scalar ratio  lies within  $\mathcal{O}(10^{-3})$  and $r < 0.07$ \citep{gb}. The scalar power spectrum for each model is taken to be $P_S = 2.43 \times 10^{-9}$.

\subsubsection*{R2 Inflation model (Starobinsky model)}
This model is based on the higher order gravitational terms with the action \citep{st}
\begin{equation}
S = \int d^4x \sqrt{-g}\frac{m_{pl}^2}{2}\left(R+\frac{R^2}{6m^2}\right),
\end{equation}
where $R$ is the Ricci scalar and $m$ is the inflaton mass.

The model can be represented in the form of Einstein gravity with a normalized inflaton field with effective potential,
\begin{equation}
V(\phi) = M^4 (1-e^{-\sqrt{2/3}\phi/m_{pl}})^2.
\end{equation}
The tensor-to-scalar ratio for this model is obtained as $r = 3.25 \times 10^{-3}$.
The slow-roll parameters  obtained for the model  are 
\begin{eqnarray}
\epsilon &=& 2.03 \times 10^{-4}, \nonumber \\
\eta &=& -1.63 \times 10^{-2}. 
\end{eqnarray}

The calculated  tensor power spectrum with the tensor spectral index $n_T = -4.06 \times 10^{-4}$  is  $P_T = 7.9 \times 10^{-12}.$
 
\subsubsection*{Arctan Inflation model}
This model is considered as a large field inflation where the inflaton field starts at a large value and then evolves to the minimum potential \citep{ai1, ai2}. The effective potential for this model is given by
\begin{equation}
V(\phi) = M^4 \left[1 - \arctan\left(\frac{\phi}{\mu}\right)\right],
\end{equation}
where $\mu/m_{pl} = 10^{-2}$ is a free parameter which characterizes the typical vacuum expectation value at which inflation takes place, $M/m_{pl} = 10^{-3}$.

The tensor-to-scalar ratio for this model is found  as $r = 1.38 \times 10^{-2}$.
The calculated slow-roll parameters are,
\begin{eqnarray}
\epsilon &=& 8.62 \times 10^{-4}, \nonumber \\
\eta &=& 3.0 \times 10^{-2}. 
\end{eqnarray}

The obtained tensor power spectrum is  $P_T = 3.35 \times 10^{-11}$ for which  the tensor spectral index has the value  $n_T = -1.72 \times 10^{-3}$.

\subsubsection*{Higgs Inflation model}
In this model, the Higgs field is considered to play the role of the inflaton. The field is considered to be non-minimally coupled to gravity\citep{higgs}. The effective potential for this model is
\begin{equation}
V(\phi) = M^4 (1 + e^{-\sqrt{2/3}\phi/m_{pl}})^{-2}.
\end{equation}
The tensor-to-scalar ratio for this model is  $r = 2.83 \times 10^{-3}$.
The corresponding slow-roll parameters are,
\begin{eqnarray}
\epsilon &=& 1.77 \times 10^{-4}, \nonumber \\
\eta &=& -1.48 \times 10^{-2}. 
\end{eqnarray}

The tensor power spectrum is  obtained as $P_T = 6.87 \times 10^{-12}$ with  $n_T = -3.53 \times 10^{-4}$.

\subsubsection*{Inverse Monomial Inflation model}
This model is considered in the context of quintessential inflation where the inflaton need not necessarily decay and hence, may survive through the present epoch. Since the inflaton does not decay, radiation is created via gravitational particle production \citep{im1, im2, im3}. The effective potential for this model is
\begin{equation}
V(\phi) = M^4 \left(\frac{\phi}{m_{pl}}\right)^{-p},
\end{equation}
where $p=3$ is a positive parameter, $M/m_{pl} = 10^{-1}$.

The calculated  tensor-to-scalar ratio for this model  is   $r = 2.0 \times 10^{-3}$ and 
the slow-roll parameters are,
\begin{eqnarray}
\epsilon &=& 1.25 \times 10^{-4}, \nonumber \\
\eta &=& 3.33 \times 10^{-4}. 
\end{eqnarray}

The tensor power spectrum is found as  $P_T = 4.86 \times 10^{-12}$ with  $n_T = -2.50 \times 10^{-4}$.

\subsubsection*{Loop Inflation model}
This model is studied in the context of spontaneous symmetry breaking which alters the flatness of the potential and takes the form of logarithmic function for one loop order correction \citep{li1, li2, li3}. The effective potential for this model is
\begin{equation}
V(\phi) = M^4 \left[1 + \alpha \ln \left(\frac{\phi}{m_{pl}}\right)\right],
\end{equation}
where $\alpha = g^2/16\pi^2=0.5$ tunes the strength of radiative effects, $M = 10^{16}$ GeV.

The tensor-to-scalar ratio for this model is calculated to be $r = 4.34 \times 10^{-2}$.
The slow-roll parameters are,
\begin{eqnarray}
\epsilon &=& 3.09 \times 10^{-3}, \nonumber \\
\eta &=& -2.06 \times 10^{-2}. 
\end{eqnarray}

The tensor power spectrum is calculated to be $P_T = 1.2 \times 10^{-10}$ with the tensor spectral index $n_T = -6.18 \times 10^{-3}$.

\section{Calculations}
Suppose horizon crossing occurs at time $t=t_{hc}$, then the critical momentum $k_{hc}$ when both the mass term and the momentum contribute equally to frequency is given by
\begin{equation}
k_{hc}=a_{hc}m_{gw}(t_{hc})=\frac{a_{hc}H_c}{\sqrt{2}}
\end{equation}
and the scale factor at re-entry time in GR is given by
\begin{equation}
a_{k0}^{GR}=\frac{k_0}{H_{k0}^{GR}}
\end{equation}
We can write Eq.\ref{eh} as a function of $k$ using Eq.\ref{jk} as
\begin{equation}
S(k) = \sqrt{2}\times 10^{-2}\left(\frac{k^2}{m_{gw}^2}\right)^{-1/2}
\end{equation}
where we have assumed $H_c \equiv H_{k0}^{GR}$, taking the values as $k_c \sim 10^{-19}$ Hz and $k_0=2 \times 10^{-18}$ Hz and dropped the subscript 0 for notational convenience.

\begin{figure}[ht]
\includegraphics[scale=0.3]{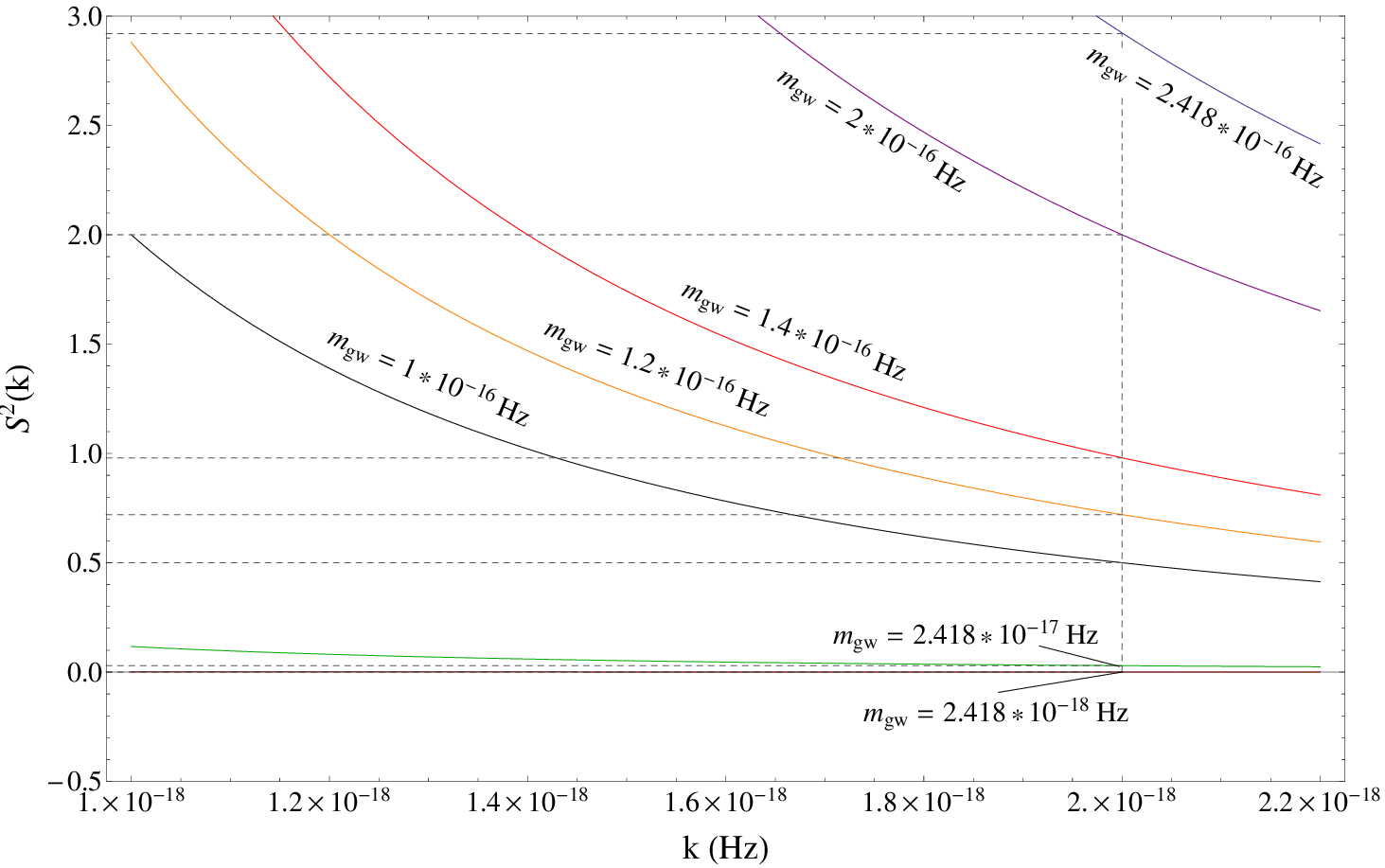}
\caption{Behavior of amplification factor $S^2(k)$ relative to $k$ for various $m_{gw}$.}\label{svk}
\end{figure}
\begin{figure}[ht]
\includegraphics[scale=0.3]{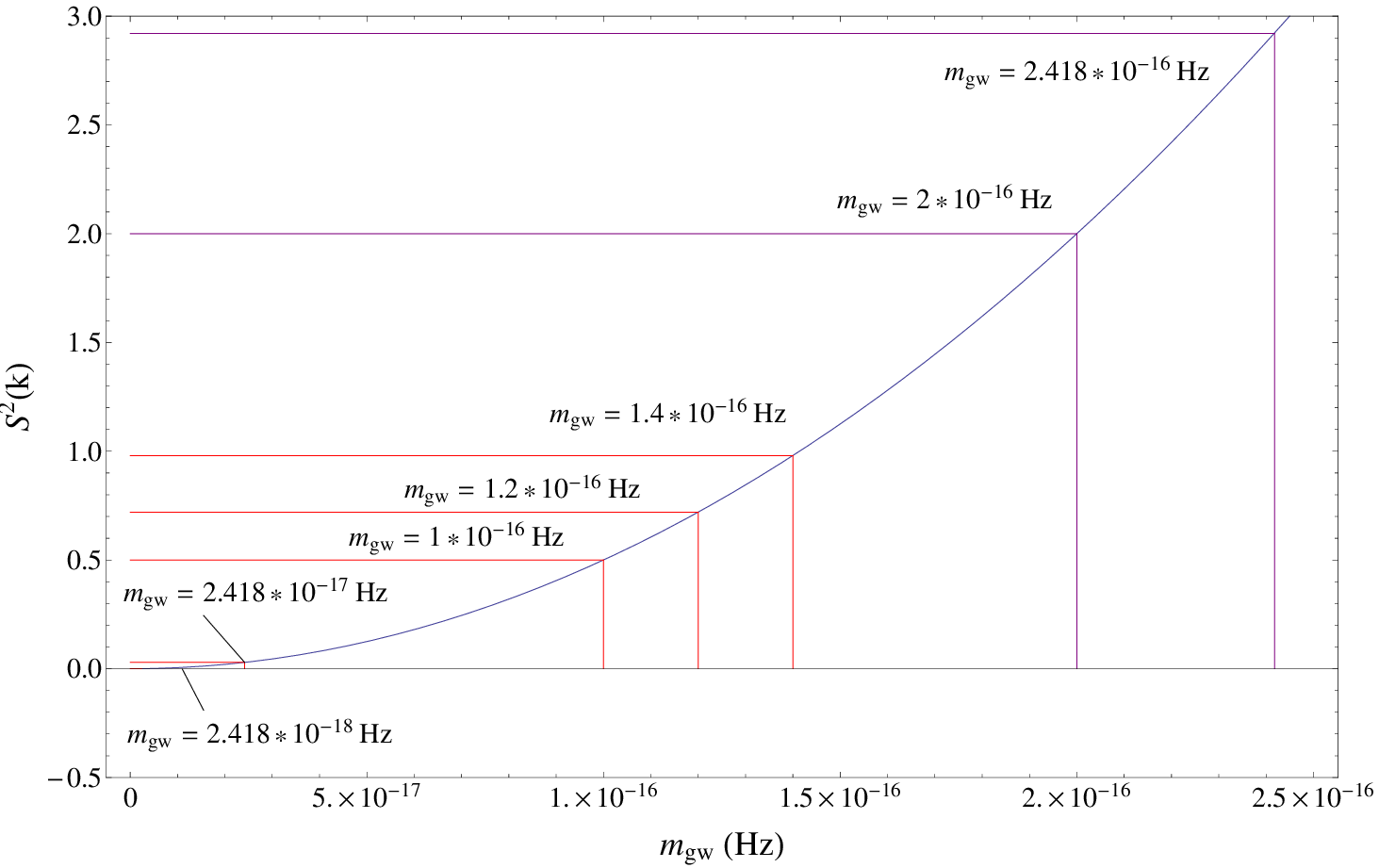}
\caption{Behavior of amplification factor $S^2(k)$ with mass $m_{gw}$. }\label{svm}
\end{figure}

It can be seen from Eq.\ref{ps} that $P_T(k) \propto S^2(k)$. In figures \ref{svk} and \ref{svm}, we show the behavior of $S^2(k)$ with $k$ and mass respectively. The wave number $k$ is very small for primordial gravitational waves in the frequency range which could produce a signature on CMB with wavelength comparable to the present-day Hubble radius. As such, the evaluation is done with the wave number comparable to the same, $k \sim 2 \times 10^{-18}$ Hz.

In our evaluation, we have taken $m_{gw} = 2.418 \times 10^{-16}~{\rm Hz} \equiv 10^{-30}$ eV as the upper bound for massive primordial gravitational waves and $m_{gw} = 2.418 \times 10^{-18}~{\rm Hz} \equiv 10^{-32}$ eV as the lower bound. For mass comparable to the inflationary Hubble scale ($\equiv 10^8$ GeV), the massive gravitons generate a blue-tilted tensor spectrum during inflation \citep{bt1,bt2}. Also, massive spin-2 particle produces a blue tilt if $-2\epsilon + 2m_{gw}^2/3H^2 > 0$ \citep{bt3}. Since the masses we have chosen are very small, it can be realized by straightforward calculations that for each model, we get red-tilted spectrum.

In figure \ref{svk}, we show the behavior of $S^2(k)$ with wave number $k$ in the long wavelength regime. The vertical dashed line indicates $k=2 \times 10^{-18}$ Hz. The horizontal dashed lines indicate the amplification factor for each mass at $k = 2 \times 10^{-18}$ Hz.

In figure \ref{svm}, the blue curve represents $k=2 \times 10^{-18}$ Hz. The purple lines indicate masses for which $S^2(k) > 1$ and the red ones for $S^2(k) < 1$. As such, masses with $S^2(k)>1$ will see enhancement in the spectrum while those with $S^2(k)<1$ will see suppression in the power level.

\section{The B-mode polarization of CMB}
The expression for computing the  $BB$-mode correlation  angular power spectrum of  CMB is \citep{cmb4, baskaran} 
\begin{eqnarray}
C_l^{BB} &=& (4\pi)^2 \int dk k^2 P_T(k) \nonumber \\ && \times \left| \int_0^{\eta_0} d\eta g(\eta) h_k(\eta) \Big\{2j'_l(x) + \frac{4j_l(x)}{x}\Big\}_{x}\right|^2
\end{eqnarray}
where $g(\eta) = \frac{d\kappa}{d\eta} e^{-\kappa}$ is the probability distribution of the last scattering with  $\kappa$ as  the differential optical depth  and $j_l(x)$ is the spherical Bessel function. The equation is evaluated at $x=k(\eta_0-\eta)$.

The CMB angular spectrum for the  BB mode correlation with the slow-roll inflation models are obtained  by using the CAMB code with  $\kappa = 0.08$  and $k_0 = 0.002$  Mpc$^{-1}$ as the tensor pivot scale. We generated the BB-mode $C_l$ data for each model using the CAMB code. Then, incorporating the massive effect, we plotted the data after adding lensing effect to the pure BB-mode. This is done so as the BKP joint data incorporates lensing effect in the errorbar. 
\begin{figure}[ht]
\includegraphics[scale=0.2]{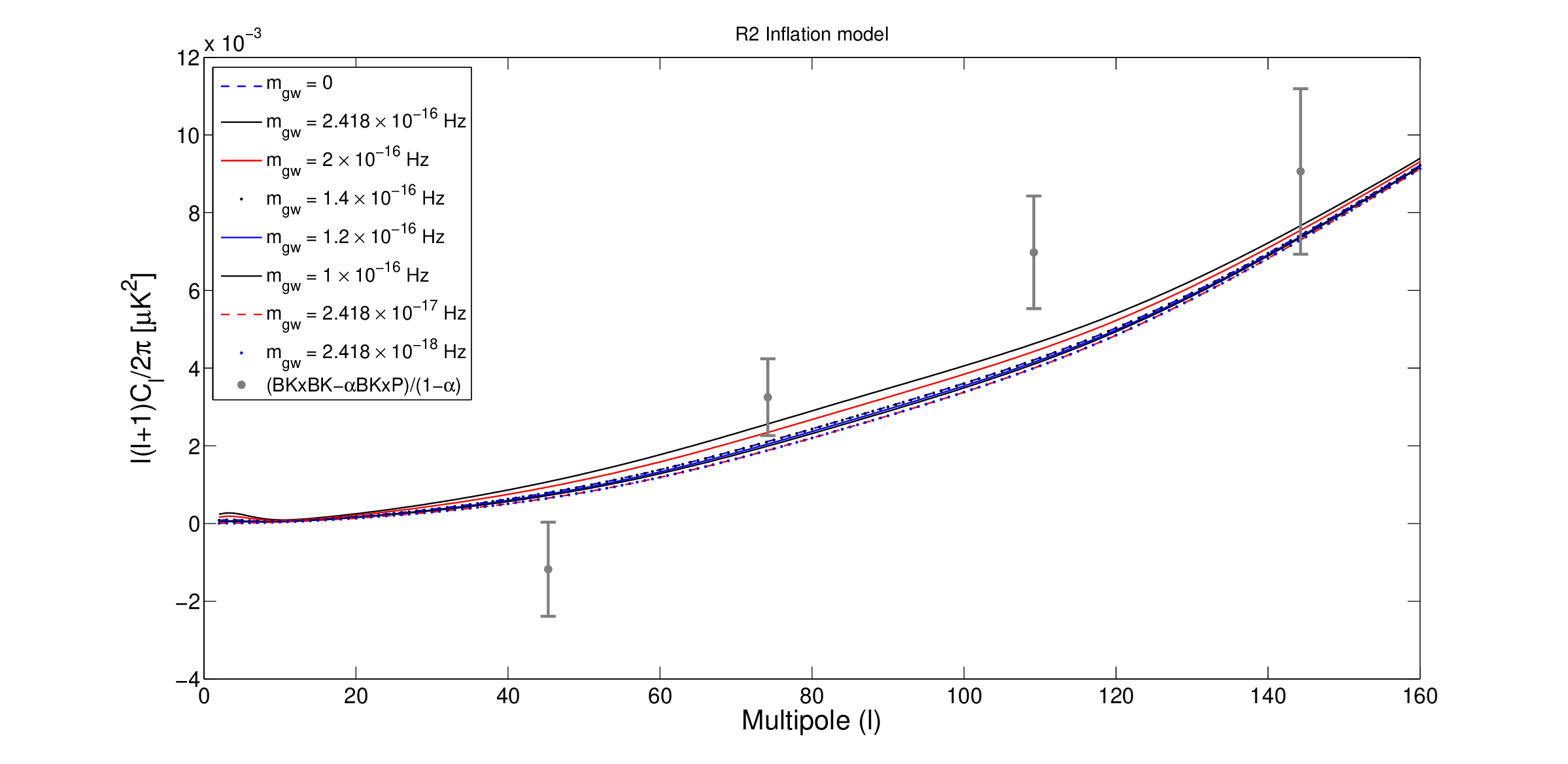}
\caption{Lensed BB-mode correlation  angular spectrum of CMB for the Starobinsky (R2) inflation model  for various values of graviton mass  with  the   BICEP2/Keck Array  and Planck  joint data.}\label{f1}
\end{figure}

\begin{figure}[ht]
\includegraphics[scale=0.2]{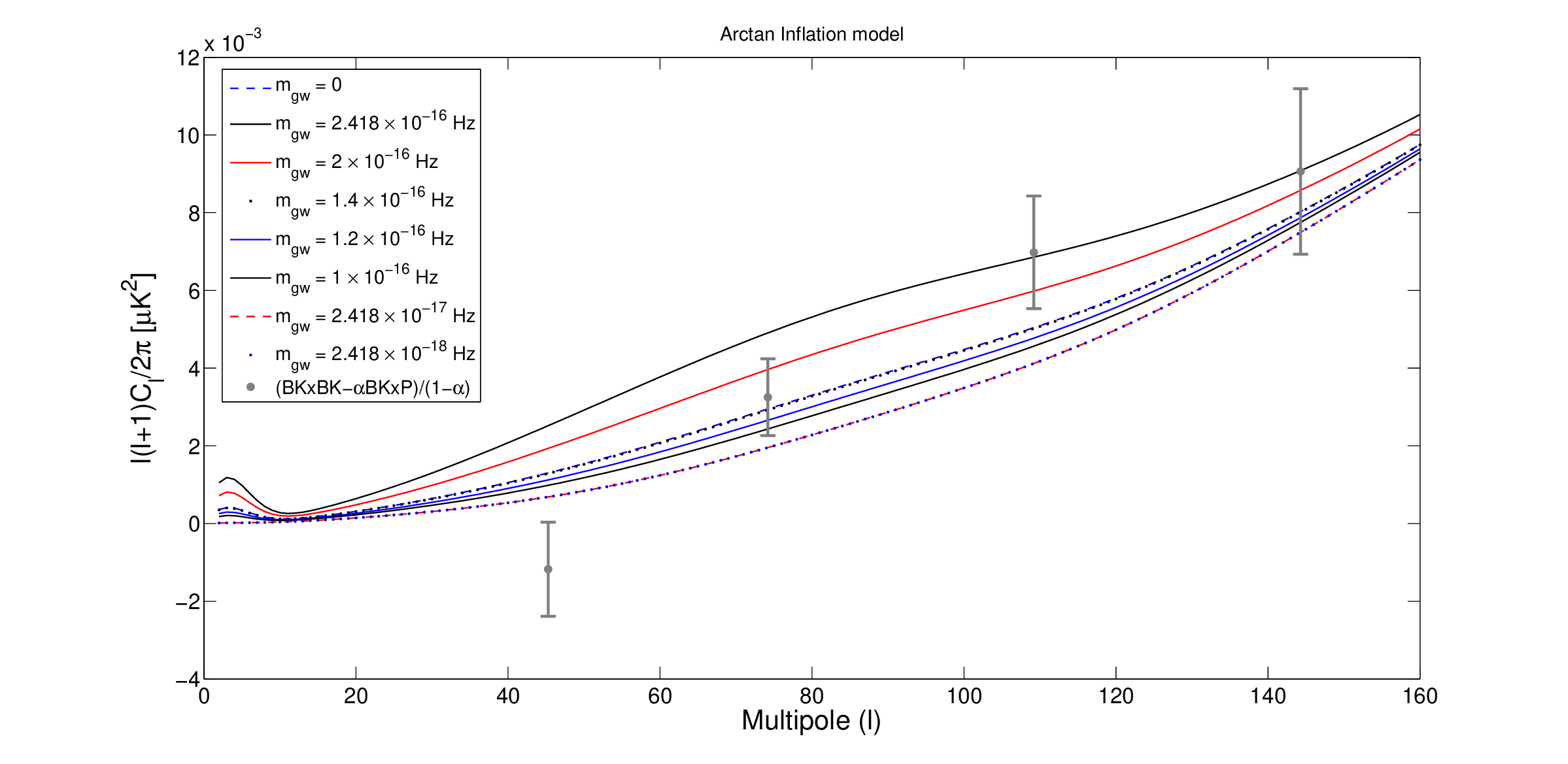}
\caption{Lensed BB-mode correlation  angular spectrum of CMB for the Arctan inflation model with with  the  recent BICEP2/Keck Array  and Planck  collaboration data lensing for various values of graviton mass with  the   BICEP2/Keck Array  and Planck  joint  data.}\label{f2}
\end{figure}

\begin{figure}[ht]
\includegraphics[scale=0.2]{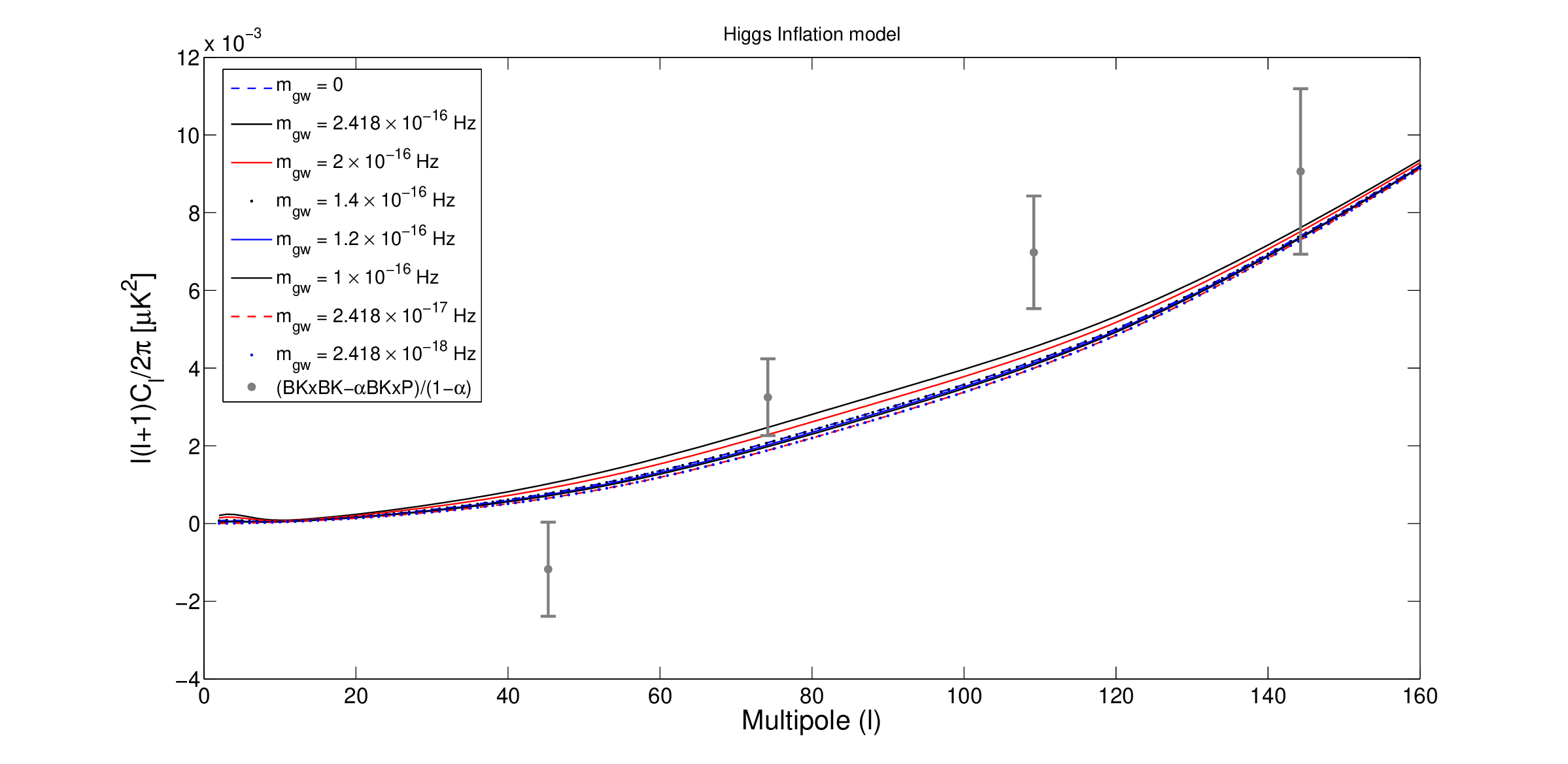}
\caption{Lensed BB-mode correlation  angular spectrum of CMB for the  Higgs inflation model  for various values of graviton mass with  the   BICEP2/Keck Array  and Planck  joint data.}\label{f3}
\end{figure}

\begin{figure}[ht]
\includegraphics[scale=0.2]{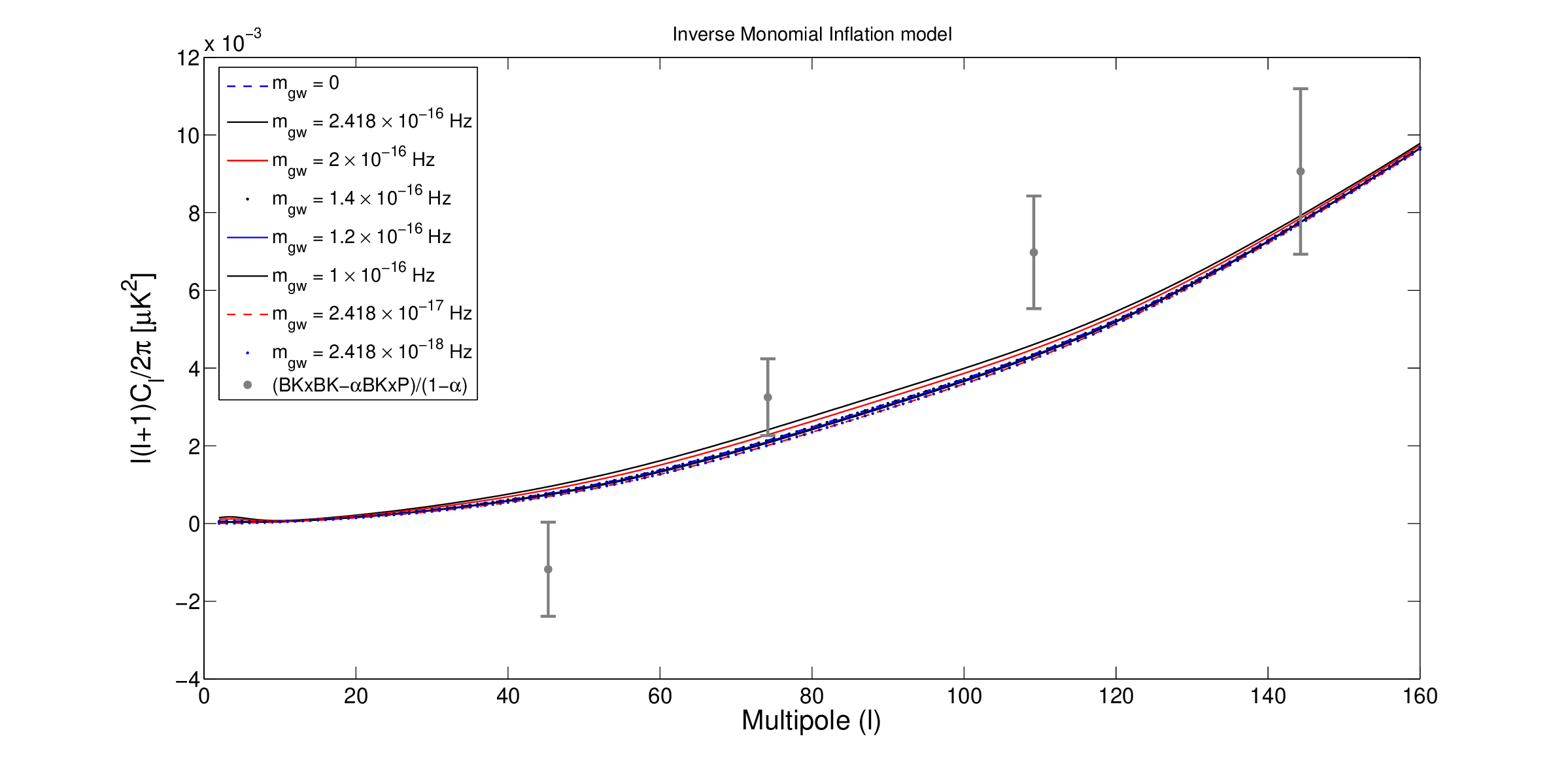}
\caption{Lensed BB-mode correlation  angular spectrum of CMB for  the Inverse monomial inflation model  for various values of graviton mass with  the   BICEP2/Keck Array  and Planck  joint  data.}\label{f4}
\end{figure}

\begin{figure}[ht]
\includegraphics[scale=0.2]{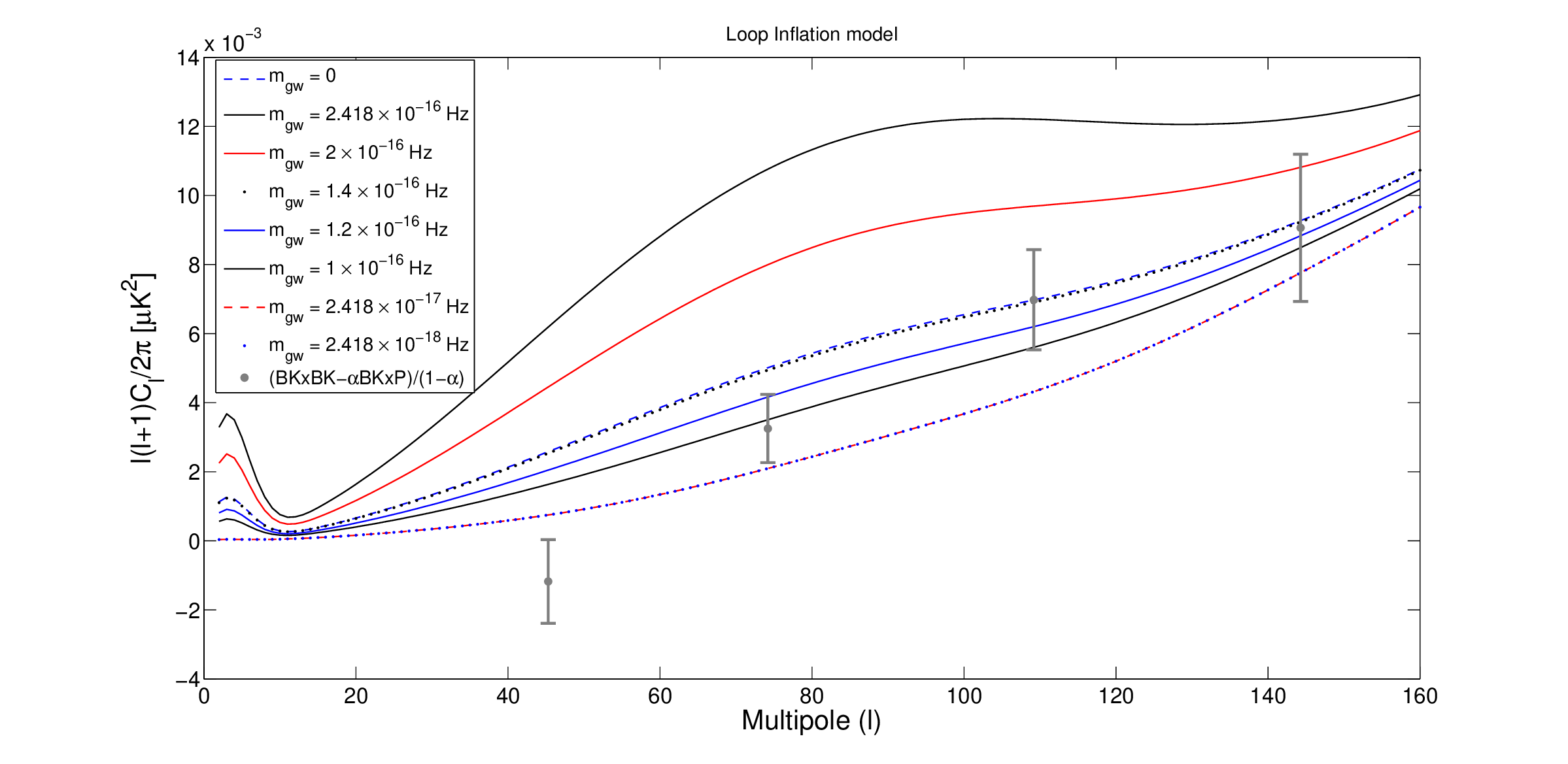}
\caption{Lensed BB-mode correlation  angular spectrum of CMB  for the Loop inflation model  for various values of graviton mass with  the   BICEP2/Keck Array  and Planck  joint data.}\label{f5}
\end{figure}
The obtained results are presented in  figures \ref{f1}, \ref{f2}, \ref{f3}, \ref{f4} and \ref{f5}.  The limit $(BK \times BK - \alpha BK \times P)/(1 - \alpha)$ is taken  from the BKP joint data after subtraction of dust contribution on the BICEP2/Keck Array band which is 4\% times more than that in the Planck band thus giving the fiducial value $\alpha = 0.04$ \citep{bkp}.

\section{Conclusion and discussion}
The BB mode correlation angular power spectrum of CMB for  the primordial massive gravitational waves   for  the Starobinsky (R2), arctan, Higgs, inverse monomial and loop  inflation models is studied in the context of Lorentz violating massive gravity model. Of the models studied, loop inflation model is marginally favored by constraints based on the BICEP2/Keck and Planck joint data while the rest are highly favored. The masses for which we have plotted the spectrum are those which have been previously proposed for primordial gravitational waves for consistency alongwith our own estimates where we have converted every mass unit into Hz \citep{Dubovsky, cmbmg2, mgtensor, hig}. Note that in the figures, the enhancement around $l \sim 80$ is more model dependent rather than mass, for instance, for models with large $r$, enhancement is more. Thus, this is relative to $r$.

It is observed for  each inflation  model that,  for gravitational waves with mass $m_{gw} \gtrsim 1.4 \times 10^{-16}$ Hz, there is enhancement in the power spectrum compared to that of the massless gravitational waves case while  there is  decrease in the   power level  in the case of  $m_{gw} < 1.4 \times 10^{-16}$ Hz.  The increase/decrease in the power level of  BB mode angular power spectrum of CMB  for the massive gravitational waves   is greater for inflation models with larger deviation ($n_T$) from scale invariance. The BB mode angular power spectrum of CMB for gravitational waves with mass $m_{gw} \simeq 1.4 \times 10^{-16}$ Hz ($\equiv 5.79 \times 10^{-31}$ eV) is found almost comparable  to  its  massless counterpart. Hence, this is the value of the mass of primordial graviton that we have obtained.

For each slow-roll inflation model, the  angular power spectrum  for  the  gravitational waves  with masses $m_{gw} = 2.418 \times 10^{-17}$ Hz ($\equiv 10^{-31}$ eV) and $m_{gw} = 2.418 \times 10^{-18}$ Hz ($\equiv 10^{-32}$ eV) are found marginally within the limit  of  BICEP2/Keck and Planck joint data at higher multipoles and well outside the limit at lower multipoles, which indicates  that the lower limit for the graviton mass may be higher than these masses. At the same time, the upper limit for the primordial graviton mass may also be higher than $m_{gw} = 10^{-30}$ eV. Hence, the results and analysis of the present study  on the  BB mode angular power spectrum of CMB with the  BICEP2/Keck Array and Planck joint data  for various inflationary models show that the mass limit for primordial graviton may be higher than  the earlier proposals.

Thus, assuming a modified dispersion relation for these waves,  the mass of the primordial graviton has been calculated and observed as $m_{gw} \approx 5.79 \times 10^{-31}$ eV at the Compton wavelength $\lambda_g = 2.1 \times 10^{21}$ km. Our resulting estimate on the mass of the graviton is also in good agreement with other theoretical estimates \citep{Dubovsky, cmbmg2, sdas}. The present study may be repeated with  other inflation models  which does not seem to alter the conclusions of the present study.

\section*{Acknowledgement}
Authors would like to thank SERB, New Delhi for financial support. N M also acknowledges financial support from CSIR, New Delhi at the beginning of this project. The authors acknowledge the use of online CAMB Tool and Bicep2/Keck Array and Planck data. The authors are also thankful to the unknown reviewers for their valuable comments, suggestions and inputs.

\section*{Conflict of interest statement}
N. Malsawmtluangi and P.K. Suresh declare that they have no conflict of interest.

\appendix

\section{Graviton mass parameters}
The quadratic Lagrangian in Eq.\ref{lm} can be written in terms of the tensor, scalar and vector fields as,
\begin{eqnarray}
L_m &=& m_{pl}^2 \Big[-\frac{1}{4}m_2^2 h_{ij}^2 - \frac{1}{2} m_2^2 (\partial_i Q_j)^2 + m_0^2 \varphi^2 + \frac{1}{2}m_1^2 (\partial_i A)^2 + (m_3^2 - m_2^2) (\partial_i^2 E)^2 \nonumber\\
  &&  - 2(3m_3^2 - m_2^2) \psi \partial_i^2 E + 3(3m_3^2 - m_2^2) \psi^2 
 + 2m_4^2 \varphi \partial_i^2 E - 6 m_4^2 \varphi \psi \Big].
\end{eqnarray}

For a particular case where the equation of state parameter $w = -(3\gamma)^{-1}$ so that $\rho_{\phi} = -3 \gamma p_{\phi}$, the mass parameters follow the relations,
\begin{eqnarray}
m_0^2 &=& 3\gamma \left(m_4^2 - \frac{m_1^2}{2}\right), \nonumber \\
m_1^2 &=& 2(3\gamma - 1) p_{\phi}, \\
m_4^2 &=& \gamma (3 m_3^2 - m_2^2). \nonumber
\end{eqnarray}

With the conditions $m_0 \neq 0$ and $m_1 \neq 0$ and $m_4 \neq 0$, there are two scalar degrees of freedom at the linear level about the flat spacetime, one of these degrees of freedom introduces a ghost mode. Hence absence of ghost mode demands either $m_0 = 0$ or $m_1 = 0$ or both $m_2 = m_3$ and $m_4 = 0$. 

When $m_0 = 0$, the scalar field $\varphi$ acts as the Lagrangian multiplier which leads to the constraint, \[ 2\partial_i \psi = m_4^2 (3\psi - \partial_i E).\] 
Thus $\psi$ remains as the only remaining dynamical scalar field and the tensor perturbation $h_{00}$ enters the action linearly. This property sufficiently ensures the ghost-free scenario.

The parameter $m_1$ is responsible for turning on a kinetic term for the scalar modes. When $m_1 = 0$, the scalar field $B$ acts as Lagrangian multiplier leading to the constraint for propagating modes as $\psi = 0$. Applying this into the massive gravity action, it can be obtained that there are no propagating modes in the scalar sector.  This property is same in the vector sector. Thus, the model is free of scalar degrees of freedom about the Minkowski at the linear level, there is no vDVZ discontinuity.

When $m_2 = m_3$ and $m_4 = 0$, the field $E$ enters the action linearly leading to the corresponding field equation, \[2 \ddot{\psi} +  (3m_3^2 - m_2^2) \psi) = 0.\] This implies the absence of high frequency propagating modes.

When the parameter $m_2^2 \geq 0$, there is no rapid instabilities in the model. In the vector sector, provided $m_2 \neq 0$, the vector field behaves in the same way as in the Einstein theory in the gauge $Q_i = 0$; hence there are no propagating vector perturbations and gravity is not modified in this sector unless one takes into account the non-linear effects or higher derivative terms.  In the scalar sector, the scalar field has massless limit which coincides with the GR expression; hence there is no vDVZ discontinuity. In the tensor sector, only the transverse-traceless perturbations $h_{ij}$ are present and their field equation is that of a massive field with the mass $m_2$ with helicity-2; hence there are two massive spin-2 propagating degrees of freedom.  Thus, this mass parameter represents the only propagating modes under the above condition which are the tensor modes, and is called the mass of the graviton.

\bibliographystyle{spr-mp-nameyear-cnd}

\end{document}